\RequirePackage{hyperref}
\documentclass[draft]{agujournal2019}
\usepackage{url} 
\usepackage{lineno}
\usepackage{soul}
\usepackage{color}
\usepackage{amsmath}
\usepackage{amsfonts}
\usepackage{float}
\usepackage{dsfont}
\usepackage[ruled,vlined]{algorithm2e}
\usepackage{tikz}
\usetikzlibrary{math,external} 
\usepackage{pgfplots} 
\pgfkeys{/pgf/images/include external/.code=\includegraphics{#1}} 
\usepackage{upgreek}

\journalname{Water Resources Research}

\begin{document}

%
%


\title{Sequential pCN-MCMC, an efficient MCMC method for Bayesian inversion of high-dimensional multi-Gaussian priors}

%
%




\authors{Sebastian Reuschen\affil{1}, Fabian Jobst\affil{1}, Wolfgang Nowak\affil{1}}


\affiliation{1}{Department of Stochastic Simulation and Safety Research for Hydrosystems, University of Stuttgart, Stuttgart, Germany}




\correspondingauthor{Sebastian Reuschen}{sebastian.reuschen@iws.uni-stuttgart.de}




\begin{keypoints}
\item A hybrid MCMC between the sequential Gibbs and the pCN-MCMC approach is presented.
\item This hybrid MCMC is more efficient than both special cases.
\item We present an adaptive algorithm for tuning the tuning-parameters of this hybrid MCMC during burn-in.
\end{keypoints}

%
%

%
%


\begin{abstract}
In geostatistics, Gaussian random fields are often used to model heterogeneities of soil or subsurface parameters. To give spatial approximations of these random fields, they are discretized. Then, different techniques of geostatistical inversion are used to condition them on measurement data. Among these techniques, Markov chain Monte Carlo (MCMC) techniques stand out, because they yield asymptotically unbiased conditional realizations. 
However, standard Markov Chain Monte Carlo (MCMC) methods suffer the curse of dimensionality when refining the discretization. This means that their efficiency decreases rapidly with an increasing number of discretization cells. Several MCMC approaches have been developed such that the MCMC efficiency does not depend on the discretization of the random field. The pre-conditioned Crank Nicolson Markov Chain Monte Carlo (pCN-MCMC) and the sequential Gibbs (or block-Gibbs) sampling are two examples. In this paper, we will present a combination of the pCN-MCMC and the sequential Gibbs sampling. Our algorithm, the sequential pCN-MCMC, will depend on two tuning-parameters: the correlation parameter $\beta$ of the pCN approach and the block size $\kappa$ of the sequential Gibbs approach. The original pCN-MCMC and the Gibbs sampling algorithm are special cases of our method. 
We present an algorithm that automatically finds the best tuning-parameter combination ($\kappa$ and $\beta$) during the burn-in-phase of the algorithm, thus choosing the best possible hybrid between the two methods. In our test cases, we achieve a speedup factors of $1-5.5$ over pCN and of $1-6.5$ over Gibbs. Furthermore, we provide the MATLAB implementation of our method as open-source code. 
\end{abstract}

%
%

\newcommand{\TODO}{\textcolor{red}{\textbf{TODO: }}}
\newcommand{\mymethod}{sequential pCN-MCMC}  

\newcommand{\pazero}{\boldsymbol{{\uptheta}}^{(0)}}
\newcommand{\pai}{\boldsymbol{{\uptheta}}^{(i)}}
\newcommand{\paii}{\boldsymbol{{\uptheta}}^{(i+1)}}
\newcommand{\paj}{\boldsymbol{{\uptheta_j}}} 
\newcommand{\param}{\boldsymbol{{\uptheta}}}
\newcommand{\paramt}{\boldsymbol{\widetilde{{\uptheta}}}}
\newcommand{\pait}{\boldsymbol{\widetilde{{\uptheta}}}^{(i)}}
\newcommand{\data}{\boldsymbol{\mathrm{d}}}

\section{Introduction}
The heterogeneity of soil parameters is a key control on subsurface flow and transport. Geostatistical methods are usually used to characterize these heterogeneities \cite<e.g.>{refsgaard2012review}. In general, all soil parameters can be described by random functions. In this work, we focus on soil parameters which can be (a priori) described by Gaussian processes. A Gaussian process is a stationary random function, in which any finite collection of variables can be described by a multivariate normal distribution. Such a distribution is fully described by a mean vector and a covariance matrix.

The goal of Bayesian inversion is to predict (and give uncertainties) of parameters given measurements. The probability distribution of parameters before measurements is called prior probability distribution, whereas the conditional probability (after measurements) is called a posterior probability distribution. If the parameters are measured directly, the Kriging (also called Gaussian process regression) procedure allows us to calculate the posterior probability distributions of all parameters analytically \cite<e.g.>{kitanidis1997}. However, if the parameters are not measured directly (here: measure the hydraulic head and infer the hydraulic conductivity), Kriging is not applicable.

Instead, sampling methods can be used to solve this problem. Examples are rejection sampling \cite<e.g>[chapter 10.2]{gelman1995bayesian}, which is applicable to low-dimensional prior distributions or weak data, Ensemble Kalman filters \cite<e.g.>{Evensen2009}, which are used to linearize forward models for multi Gaussian posteriors, and more. Here, we will focus on Markov chain Monte Carlo (MCMC) methods which are universally applicable for Bayesian inference \cite<e.g.>{QIAN2003269} but computationally expensive. 

In MCMC approaches, the random function is discretized to enable numerical computations. The main problem of most common MCMC methods, e.g. the Metropolis-Hastings algorithm \cite{metropolis1953,Hastings1970}, is, that a refinement of discretization leads to worse convergence speed of the methods \cite{Cotter2013}. Different approaches have been presented in the literature to overcome this challenge. In the following, we present two approaches.

The key idea of the (sequential) Gibbs approach \cite<e.g.>[chapter 11.3]{gelman1995bayesian} is to sample one (or several) parameters conditionally, with respect to the prior distribution, on all remaining parameters. This leads to a discretization-independent efficiency for arbitrary prior distributions. 
The limitations of the Gibbs approach are that the conditional sampling from the prior needs to be possible and computationally cheap \cite<e.g.>{Fu2008}. In most applications, however, the forward simulation is the computational bottleneck.
One specific version of this approach was proposed by \citeA{Fu2008,Fu2009a,Fu2009b}, who resampled boxes of the multi Gaussian parameter field. \citeA{Hansen2012} applied this idea to resample boxes of the parameter field in a binary classification problem. A combination of these approaches for binary classification problems with multi-Gaussian heterogeneity was presented by \citeA{reuschen2020}.

The second discretization-independent approach we discuss is the pre-conditioned Crank Nicolson MCMC (pCN-MCMC) \cite{Beskos2008,Cotter2013}. It is easy to implement and computationally fast, as demonstrated in the respective original papers and in a recent effort to construct reference solutions algorithms for geostatistical inversion benchmarks \cite<e.g.>{Xu2020}. In fact, pCN-MCMC has been derived for inverting random \emph{functions}, so that the numerical discretization of the random field does not matter for its convergence speed by construction. However, the pCN-MCMC can only be used for multi-Gaussian priors. The reason for this restriction is that the pCN proposed modifications to random fieldy by a small-magniture random field to a dampened version of the current field, thus resembling an autoregressive process of order one along the chain \cite{Beskos2008}. While this way of proposing new solutions is highly effective and independent in convergence speed of the spatial discretization, it can only be constructed when the prior is multi-Gaussian. 

Other alternatives using spectral parametrization \cite{Laloy2015}, Karhunen-Loeve expansions \cite<e.g.>{Mondal2014} or pilot point methods \cite<e.g.>{Jardani2013} use dimension reduction approaches for fast convergence. The consequences of these approaches are twofold:
On the one hand, dimension reduction approaches can reduce computational cost. On the other hand, they only converge towards approximate solutions of the true posterior. In this work, we focus on methods that converge to the true posterior.

Another direction of research uses derivatives of the posterior distribution to increase the efficiency of MCMC methods \cite<e.g. Hamiltonian MCMC as summarized in >{betancourt2017conceptual}. Analytical derivatives of the likelihood function are possible in some scenarios, but in most hydraulic forward models, they are impossible.  Numerical approximation of gradients is an alternative. However, numerical differentiation is computationally expensive and often negates the advantage of these methods. Hence, we focus on methods that do not require gradient information. 

We combine the sequential Gibbs idea with the pCN-MCMC idea and create a hybrid method called sequential pCN-MCMC. However, the hybrid method comes with two tuning-parameters. The standard way of finding the optimal tuning-parameters in MCMC algorithms for high-dimensional inverse problems is to tune them for an acceptance rate equal to 23.4\% \cite<see>{Gelman1996}. In our hybrid method, this is not possible anymore because infinitely many tuning-parameter combinations lead to the same acceptance rate. Hence, we refrain to the efficiency defined in \citeA{Gelman1996} to find the optimal tuning-parameter combination.

Overall, the novelty of our paper is a combination of the sequential Gibbs MCMC and the pCN-MCMC, which we call sequential pCN-MCMC. Here, the pCN-MCMC and sequential Gibbs are special cases of the sequential pCN-MCMC. Our hypothesis is that the most efficient method is neither of the special cases.

We compare the new hybrid method to the two original algorithms for Bayesian inversion of fully saturated groundwater flow. Here, we use different scenarios where we alter the prior information and discretization to test and confirm our hypothesis. 
The MATLAB implementation of our code is available at \href{https://bitbucket.org/Reuschen/sequential-pcn-mcmc}{https://bitbucket.org/Reuschen/sequential-pcn-mcmc}.

The paper is structured as follows. Section \ref{chp_methods} gives a definition of the inverse problem, an overview over existing methods and introduces metrics to evaluate the performance of algorithms. In section \ref{chp_my_method}, we present our proposed sequential pCN-MCMC method.  After that, we introduce our test cases in section \ref{chp_test_cases}. Our results are shown in section \ref{chp_results} and discussed in section \ref{chp_discussion}. Finally, section \ref{chp_conclusion} concludes the most important findings in a short summary.

\section{Methods}
\label{chp_methods}
In this section, we briefly recall existing MCMC methods for multi-Gaussian priors. Again, we focus on those without dimensionality reduction and without derivatives. First, we give the definition of the problem class in section \ref{chp_bayesian}. In section \ref{chp_MCMC}, we introduce the generic MCMC approach. After that, we recall the Metropolis-Hastings approach in section \ref{chp_metropolis} and discuss the differences to the so-called prior sampling methods in section \ref{chp_prior_sampling}. Sections \ref{chp_pcn} and \ref{chp_gibbs} introduce the existing algorithms pCN-MCMC and sequential Gibbs sampling, respectively, which are both instances of prior sampling methods. Finally, we present metrics to evaluate the presented methods in section \ref{chp_metrics}.

\subsection{Bayesian inference}
\label{chp_bayesian}
\noindent Let 
\begin{align}
\label{eqn_bayesian_inv}
\data = F(\boldsymbol{\mathrm{\Theta}})  + \boldsymbol{\mathrm{e}}
\end{align}
be the stochastic representation of a forward problem. $F(\boldsymbol{\Theta})$ is an error-free deterministic forward model. Equation \ref{eqn_bayesian_inv} describes the relation between the unknown and uncertain parameters $\boldsymbol{\Theta}$ and the measurements $\data$. The noise term $\boldsymbol{\mathrm{e}}$ aggregates all error terms. The goal of Bayesian inversion is to infer the posterior parameter distribution of $\boldsymbol{\Theta}$ based on prior knowledge of $\boldsymbol{\Theta}$ and the data $\data$ under the model $F$.\\
We use the parameters $\param$  to refer to realizations of the random variable $\boldsymbol{\Theta}$ with some prior distribution $p(\boldsymbol{\Theta})$ and a posterior distribution $p(\boldsymbol{\Theta}|\boldsymbol{\mathrm{d}})$. The resulting posterior density can be evaluated for each realization $\param$ as 
\begin{align}
\label{eqn_posterior_formula}
p(\param|\data) = \frac{p(\param) p(\data|\param)}{p(\data)} \propto p(\param) p(\data|\param) = P(\param) L(\param|\data) \ .
\end{align}
In this paper, we define the prior distribution $P(\param) := p(\param)$ and the likelihood $L(\param) := p(\data|\param)$ for shorter notation. This likelihood assumes that the data $\data$ do not change during the run-time of the algorithm. 
The challenge in high-dimensional Bayesian inversion is to sample efficiently from the posterior distribution $p(\param|\data)$.

\subsection{Generic Markov Chain Monte Carlo}
\label{chp_MCMC}
\noindent Markov Chain Monte Carlo (MCMC) is a popular, accurate, but typically time-intensive algorithm to sample from the posterior distribution. In contrast to other methods, it only needs the un-normalized posterior density 

\begin{align}
\label{eqn_Prior_Likelihood}
\pi(\param) = P(\param) L(\param) \propto p(\param|\data)  .
\end{align}
to sample from the posterior distribution $p(\param|\data)$.

In the following, we name all properties that an MCMC method needs to fulfill to converge to the exact posterior distribution. Based on these, we derive the formulas of our proposed MCMC.  A general introduction to MCMC can be found in \citeA{Chib1995}. 

MCMC methods converge to $\pi$ (as presented equation \ref{eqn_Prior_Likelihood}) at the limit of infinite runtime \cite{Smith1993} if and only if irreducibility, aperiodicity and the detailed balance are fulfilled. Irreducibility and aperiodicity are fulfilled for multi-Gaussian proposals (see below) in continuous problems, which are typically used for engineering purposes. Consequently, we focus on the detailed balance in the following. Given any two parameter sets $\param$ and $\paramt$, the detailed balance is defined as

\begin{align}
\label{eqn_derivation1}
\pi(\param) h(\param,\paramt) = \pi(\paramt) h(\paramt,\param)
\end{align}
with the transition kernel $h$, which is defined as

\begin{align}
\label{eqn_derivation2}
h(\param,\paramt) = q(\param,\paramt)\alpha(\param,\paramt) \ .
\end{align}
The term $q(\param,\paramt)$ refers to the proposal distribution and $\alpha(\param,\paramt)$ to the acceptance probability. Here, $\paramt$ is proposed based on the current parameter $\param$. Equations \ref{eqn_Prior_Likelihood},  \ref{eqn_derivation1} and \ref{eqn_derivation2} can be combined to 

\begin{align}
\label{eqn_alpha}
\alpha(\param,\paramt) = min \left[\frac{P(\paramt)L(\paramt) q(\paramt,\param)}{P(\param)L(\param) q(\param,\paramt)} ,1    \right].
\end{align}

Equation \ref{eqn_alpha} provides an $\alpha$ such that the detailed balance is fulfilled. This holds for any prior $P$, any likelihood $L$ and any proposal distribution $q$.
Consequently, an infinite number of possible proposal distributions $q$  exist \cite[chapter 11.5]{gelman1995bayesian} .
This raises the questions of how to choose $q$ for fast convergence in a given problem class.

Fast convergence(after burn-in) is mainly a question of low autocorrelation of successive samples \cite{Gelman1996}. This is achieved by large changes in the parameter space. As a result, it is desirable to propose far jumps for new candidate points $\paramt$ via the proposal function $q$ and still hope to accept them with a high probability $\alpha$. However, in most practical cases, these two properties contradict each other: Making large changes in $\param$ results in distinct $P(\paramt)L(\paramt)$ and $P(\param)L(\param)$, which results in a small $\alpha$. In contrast, small changes in $\param$ results in similar $P(\paramt)L(\paramt)$ and $P(\param)L(\param)$ (if the prior and the likelihood function are smooth), which results in $\alpha$ close to 1. Thus, \citeA{Gelman1996} stated that a trade-off between the size of the change and the acceptance rate needs to be found.

\subsection{Metropolis Hastings}
\label{chp_metropolis}
\noindent The  Metropolis-Hastings (MH) algorithm \cite{metropolis1953,Hastings1970} can be used with arbitrary proposal functions. Here, we present the random walk MH algorithm. It assumes a symmetric proposal distribution

\begin{align}
\label{eqn_metropolis_hastings}
q(\param,\paramt) =  q(\paramt,\param) \ .
\end{align}
Inserting this into equation \ref{eqn_alpha}, it follows that

\begin{align}
\alpha(\param,\paramt) = \text{min} \left[\frac{P(\paramt)L(\paramt)}{P(\param)L(\param)} ,1    \right]  = \text{min} \left[\frac{\pi(\paramt)}{\pi(\param)} ,1    \right].
\end{align}
The MH algorithm samples from any parameter distribution $\pi(\param)$. 
The specific proposal function $q$ of the so-called random walk MH is given by 

\begin{align}
\label{eqn_MH_prop}
\paramt = \param + \beta \boldsymbol{\upxi},\quad \boldsymbol{\upxi} \sim N(0,\mathds{1}) \ .
\end{align}
Here, the parameter $\beta$ controls how big the change between successive parameters $\param$ is. The proposal function $q$ of the random walk MH algorithm fullfils equation \ref{eqn_metropolis_hastings} because the proposal step (equation \ref{eqn_MH_prop}) is symmetric per definition. 

The main weakness of the Metropolis-Hastings algorithm is that the acceptance rate in equation \ref{eqn_metropolis_hastings} decreases rapidly for increasing $\beta$, especially in high-dimensional problems \cite{Roberts2002}. 
This can be improved by using the additional information that $\pi(\param) = P(\param) L(\param)$.
This enables us to make the acceptance rate $\alpha$ only dependent on $L(\param)$ in the next section.

\subsection{Prior sampling}
\label{chp_prior_sampling}
\noindent Bayesian inversion methods often exploit the knowledge that the posterior distribution follows by construction from  $\pi(\param) = P(\param)L(\param)$ to increase the efficiency (see section \ref{chp_efficiency} for the definition of efficiency).
To exploit this situation, the a-priori knowledge contained in the prior distribution $P(\param)$ is used to define a tailored proposal distribution  $q(\param,\paramt)$ (which is only efficient for the respective prior distribution). 
Mathematically, this is realized by defining $q(\param,\paramt)$ such that it fulfills

\begin{align}
\label{eqn_proposal}
q(\param,\paramt) = \frac{P(\paramt)}{P(\param)} q(\paramt,\param) . 
\end{align}

Combining equation \ref{eqn_alpha} and  \ref{eqn_proposal} results in \cite<e.g.>[]{Tarantola2005}

\begin{align}
\label{eqn_proposal_alpha}
\alpha(\param,\paramt) = \text{min} \left[\frac{L(\paramt)}{L(\param)} ,1    \right] .
\end{align}

Many problem classes, e.g. high-dimensional geoscience problems, have a complex prior $P(\param)$. 
As a result, the acceptance rate $\alpha$ is almost exclusively dependent on the prior. This leads to decreasing efficiencies of the MCMC \cite{Roberts2002}. Equations \ref{eqn_proposal} and \ref{eqn_proposal_alpha} enable us to circumvent that problem by making the acceptance rate $\alpha$ only dependent on the likelihood, because the prior is already considered in the proposal distribution. This makes it possible to have high acceptance rates $\alpha$ even for far jumps in the parameter space, which is synonymous with a high efficiency (see section \ref{chp_efficiency}). We call this approach "sampling from the prior distribution" \cite{reuschen2020}.

In the following, we will present the preconditioned Crank Nicolson MCMC (pCN-MCMC) and the block Gibbs MCMC algorithms. The proposal functions of both methods fulfill equation \ref{eqn_proposal}. Based on them, we propose our new \mymethod\ algorithm, which combines the approaches of pCN and Gibbs.

\subsection{pCN-MCMC}
\label{chp_pcn}
The idea of the preconditioned Crank Nicolson MCMC (pCN-MCMC) was first introduced by \citeA{Beskos2008}, who called it a Langevin MCMC approach. In 2013, \citeA{Cotter2013} revived the idea and named it pCN-MCMC.\\
The pCN-MCMC takes the assumption that the prior $P(\param)$ is multi Gaussian ($\param \sim N(\boldsymbol{\upmu},\boldsymbol{\Sigma})$). For these priors, the proposal step of the pCN-MCMC 
\begin{align}
\label{eqn_pcn}
\boldsymbol{\widetilde{\uptheta}}^{(i)} = \sqrt{(1-\beta^2)}\left(\pai -  \boldsymbol{\upmu} \right) + \beta  \boldsymbol{\upxi}^{(i)} + \boldsymbol{\upmu} , \quad \boldsymbol{\upxi}^{(i)} \sim N(\boldsymbol{0},{\boldsymbol{\Sigma}})
\end{align}
fulfills equation \ref{eqn_proposal}. Hence, the acceptance probability $\alpha$ is only dependend on the likelihood as denoted in equation \ref{eqn_alpha}. 
The tuning-parameter $\beta$ of the pCN-MCMC specifies the change between subsequent samples. For $\beta = 1$, subsequent samples are independent of each other. For lower $\beta$, the similarity of samples increases up to the theoretical limit of $\beta =0$ where subsequent samples are identical. 
In most applications, similar samples lead to similar likelihoods and a high acceptance rate. Hence, the tuning-parameter $\beta$ can be used to adjust the acceptance rate of pCN-MCMC algorithms (large $\beta$ lead to low acceptance rates and vice versa).
A pseudo code of the pCN-MCMC is presented in the Appendix.

\subsection{Sequential Gibbs sampling}
\label{chp_gibbs}
In 1987, \citeA{geman1987stochastic} introduced Gibbs sampling as a specific instance of equation \ref{eqn_proposal}. The basic concept of Gibbs sampling is to resample parts of the parameter space $\param$. In the geostatistical context, this typically means to select a random box within the parameter field, and then to generate a new random field within that box while keeping the parts outside the box fixed. The new random part is sampled from the prior, but under the condition that it must match (e.g. by conditional sampling) with the outside part. For illustrative examples on Gibbs sampling, we refer to \cite{gelman1995bayesian}.

Assuming a random parameter vector $\param$ of size $N_p \times 1$ ($N_p$ denotes number of parameters) and some permutation matrix $\boldsymbol{\mathrm{M}}$ (usually called $P_\pi$ in the literature), we can order the random variables into two parts

\begin{align}
\label{eqn_M}
\left[
\begin{array}{c}
\param_1\\
\param_2\\
\end{array}
\right] 
=
\boldsymbol{\mathrm{M}} \param \text{ with size }
\left[
\begin{array}{c}
q \times 1\\
(N_p - q) \times 1\\
\end{array}
\right] ,
\end{align}
where $\param_1$ incorporates all parameters which will be resampled conditionally on $\param_2$. The number of resampled parameters is given by $q$. A new proposal is defined as

\begin{align}
\label{eqn_proposal_general}
\paramt = \boldsymbol{\mathrm{M}}^{-1} \left[
\begin{array}{c}
\paramt_1\\
\paramt_2\\
\end{array}
\right]  = 
\boldsymbol{\mathrm{M}}^{T} \left[
\begin{array}{c}
\paramt_1\\
\paramt_2\\
\end{array}
\right]  = \boldsymbol{\mathrm{M}}^{T} 
\left[
\begin{array}{c}
\boldsymbol{\upxi}\\
\param_2 = \boldsymbol{\mathrm{r}}\\
\end{array}
\right]  , \quad \boldsymbol{\upxi} \sim p_\pi(\param_1|\param_2 = \boldsymbol{\mathrm{r}}) \ .
\end{align}
Here, the values $\boldsymbol{\mathrm{r}}$ of $\param_2$ remain constant, whereas the first part of the parameter space $\param_1$ gets resampled conditional on  $\param_2$. This approach is applicable to any propability distribution for which the conditional probability distribution $ p_\pi(\param_1|\param_2 = \boldsymbol{\mathrm{r}})$ can be sampled.

In this work, we follow the approach of \citeA{Fu2008,Fu2009a,Fu2009b} to resample boxes in a parameter space representing a two-dimensional domain.
Let $\boldsymbol{\uptheta}$ be a discretization of some parameter field (e.g. hydraulic conductivity). Here $\boldsymbol{\uptheta}(x,y)$ is the value of the parameter $\boldsymbol{\uptheta}$ at the spatial position $(x,y) \in ([0,l_x],[0,l_y])$. Hereby, let $l_x$ and $l_y$ be the length of the investigated domain in $x$ and $y$ direction. To determine $M$, we use a parameter and $\kappa \in (0,1]$ that defines the size of the resampled box as defined in equation \ref{eqn_box}, where a larger $\kappa$ corresponds to a larger resampling-box. To include the dependence of $\boldsymbol{\mathrm{M}}$ on $\kappa$, we will denote it as $\boldsymbol{\mathrm{M}}_\kappa$ in the following.

Based on a randomly chosen center point $(x^*,y^*)$ (which is re-chosen every MCMC step), we can choose  $\boldsymbol{\mathrm{M}}_\kappa$ such that

\begin{align}
\label{eqn_box}
\left[
\begin{array}{c}
\boldsymbol{\uptheta}_1\\
\boldsymbol{\uptheta}_2\\
\end{array}
\right]
= \left[
\begin{array}{l}
\{\boldsymbol{\uptheta}(x,y)\} \text{, with } |\frac{x-x^*}{l_x}|\leq\kappa \text{ and } |\frac{y-y^*}{l_y}|\leq\kappa \\
\{\boldsymbol{\uptheta}(x,y)\} \text{, with } |\frac{x-x^*}{l_x}|>\kappa   \text{ or } |\frac{y-y^*}{l_y}|>\kappa \\
\end{array}
\right] = \boldsymbol{\mathrm{M}}_\kappa \param \ .
\end{align}
This means that all parameters $\param(x,y)$ with a distance smaller than $\kappa$ to the centerpoint $(x,y)$ are part of the parameter set $\boldsymbol{\uptheta}_1$ and all $\param(x,y)$ with a distance larger than $\kappa$ are part of the parameter set $\boldsymbol{\uptheta}_2$. Pseudo code for computing $\boldsymbol{\mathrm{M}}_\kappa$ is shown in the Appendix.

Following \citeA{Fu2008,Fu2009a,Fu2009b}, this work will focus on multi-Gaussian priors.
In a multi-Gaussian prior setting,  the prior probability distribution is only based on the mean vector $\boldsymbol{\upmu}$ and covariance matrix $\boldsymbol{\Sigma}$.
According to equation \ref{eqn_M}, we portion $\boldsymbol{\upmu}$ and $\boldsymbol{\Sigma}$ as follows

\begin{align}
\boldsymbol{\upmu} = \left[\begin{array}{c}
\boldsymbol{\upmu_1} \\ \boldsymbol{\upmu_2}
\end{array}\right] \text{ with sizes }
\left[
\begin{array}{c}
q \times 1\\
(N_p - q) \times 1\\
\end{array}
\right] \ ,
\end{align}
\begin{align}
\boldsymbol{\Sigma} = \left[\begin{array}{cc}
\boldsymbol{\Sigma}_{11} & \boldsymbol{\Sigma}_{12} \\ \boldsymbol{\Sigma}_{21} & \boldsymbol{\Sigma}_{22}
\end{array}\right] \text{ with sizes }
\left[
\begin{array}{cc}
q \times q    & q \times (N_p - q) \\
(N_p - q) \times q   & (N_p - q) \times (N_p - q)\\
\end{array}
\right] \ .
\end{align}

With that, we can express the resampled parameter distribution for $\boldsymbol{\widetilde{\uptheta}}_1$ as $P(\boldsymbol{\widetilde{\uptheta}}_1|\param_2 = \boldsymbol{\mathrm{r}}) \sim N(\widetilde{\boldsymbol{\upmu}_1},\widetilde{\boldsymbol{\Sigma}}_{11})$.

The Kriging (or Gaussian progress regression) theory \cite<e.g.>{Rasmussen2006} states that the conditional probability is multi-Gaussian with mean

\begin{align}
\label{eqn_gibbs_mu}
\widetilde{\boldsymbol{\upmu}}_1 = \boldsymbol{\upmu}_1 + \boldsymbol{\Sigma}_{12}\boldsymbol{\Sigma}_{22}^{-1}(\boldsymbol{\mathrm{r}} -\boldsymbol{\upmu}_2)
\end{align}
and the covariance matrix

\begin{align}
\label{eqn_gibbs_covariance}
\widetilde{\boldsymbol{\Sigma}}_{11}= \boldsymbol{\Sigma}_{11} - \boldsymbol{\Sigma}_{12}\boldsymbol{\Sigma}_{22}^{-1}\boldsymbol{\Sigma}_{21} \ .
\end{align}
After combining equation \ref{eqn_proposal_general}, equation \ref{eqn_gibbs_mu} and \ref{eqn_gibbs_covariance}, we arrive at the proposal distribution

\begin{align}
\label{eqn_gibbs_proposal}
\paramt = \boldsymbol{\mathrm{M}}^{T} \left[\begin{matrix}
\paramt_1 \\
\param_2 \\
\end{matrix}\right] , \quad \paramt_1 \sim N(\widetilde{\boldsymbol{\upmu}}_1,\widetilde{\boldsymbol{\Sigma}}_{11}) 
\end{align}
which fulfills equation \ref{eqn_proposal}. 

The tuning-parameter $\kappa$ specifies the size of the resampling box and therefore the change between subsequent samples. Thereby, smaller $\kappa$ will lead to more similarity of subsequent samples and hence, to higher acceptance rates. 
The Appendix includes a pseudo code of the sequential Gibbs sampling method.

\subsection{Metrics}
\label{chp_metrics}
The quality of MCMC methods can be quantified using different metrics. An overview over such metrics can be found in \citeA{cowles1996} or \citeA{Roy2020}.
We use the following four test metrics.

\subsubsection{Acceptance rate $\overline{\alpha}$}
\label{chp_alpha}
The acceptance rate $\overline{\alpha}$ is the fraction of proposals that get accepted divided by the total number of proposals. \citeA{Gelman1996} showed empirically that $\overline{\alpha} =0.234$ is optimal for normal target distributions. This value of $\overline{\alpha}$ is often used to optimize the tuning-parameter (e.g. $\beta$) of MCMC runs because it is easy to implement. 

\subsubsection{Efficiency}
\label{chp_efficiency}
The efficiency of one parameter $j$ within a MCMC chain is defined as \cite<e.g.>{Gelman1996}

\begin{align}
eff_j = \frac{1}{1+2 \cdot \sum_{i=1}^{\inf} \rho_i  }
\end{align}
where $\rho_i$ is the autocorrelation of the chain with lag $i$. With this, the effective sample size ($ESS$) \cite{robert2013monte} is defined as

\begin{align}
ESS_j = eff_j \cdot N
\end{align}
with $N$ being the total number of MCMC samples. The $ESS$ represents the number of independent samples equivalent (i.e. having the same error) to a set of correlated MCMC samples. Hence, the efficiency (or $ESS$) can be used to estimate the number of MCMC samples needed to get a certain number of independent samples.  

In the following, we aggregate the individual efficiencies of all parameters to one combined efficiency. Therefore, we define the efficiency of several parameters as 

\begin{align}
eff = \frac{1}{1+2 \cdot \frac{1}{N_{p}} \sum_{j=1}^{N_{p}} \sum_{i=1}^{\inf} \rho_{i,j}  }
\end{align}
with $N_{p}$ being the number of parameters and $\rho_{i,j}$ being the autocorrelation of length $i$ of the $j$-th parameter.

\subsubsection{R-statistic}
The potential scale reduction factor $\sqrt{\widehat{R}}$ introduced by \citeA{Gelman1992} is a popular method for MCMC diagnostics. 
It measures the similarity of posterior distributions, generated by different independent MCMC chains, by comparing their first two moments.
Similarity between posterior distributions suggests convergence of the chains. This enables a convergence test in the absence of reference solutions. \citeA{gelman1995bayesian} stated that $\sqrt{\widehat{R}} \leq 1.2$ signifies acceptable convergence. 

\subsubsection{Kullback-Leibler divergence} 
The Kullback-Leibler divergence \cite{Kullback1951} is a measure to compare probability density functions. In this paper, we estimate the marginal density of each parameter and compute the KL-divergence of the MCMC chain to a reference solution. This leads us to as many KL divergences as we have parameters. To aggregate the KL divergences over all parameters, we only report the mean value.

The KL-divergence is used solely as a post-processing metric in our work. 
The advantage in comparison to the R-statistic is twofold in our case: First, we use the R-statistic to show that two independent chains converge to the same distribution. In contrast, we use the KL-divergence to show convergence to a previously calculated reference distribution. Second, the R-statistic shows convergence in the first two moment whereas the KL-divergence shows convergence in the entire distribution.

\section{Sequential pCN-MCMC}
\label{chp_my_method}
\begin{figure}
	\centering
	\begin{tikzpicture}
	
	\node [above right] at (3,0) {
%
%
\begin{tikzpicture}

\begin{axis}[%
width=1.019in,
height=1.019in,
at={(2.343in,3.634in)},
scale only axis,
point meta min=-3,
point meta max=3,
axis on top,
xmin=0.5,
xmax=50.5,
xtick={\empty},
ymin=0.5,
ymax=50.5,
ytick={\empty},
axis background/.style={fill=white},
legend style={legend cell align=left, align=left, draw=white!15!black},
colormap/jet,
colorbar right,
colorbar style={anchor=south west, at={(2.716,0)}, height=1*\pgfkeysvalueof{/pgfplots/parent axis height}}
]
\addplot [forget plot] graphics [xmin=0.5, xmax=50.5, ymin=0.5, ymax=50.5] {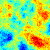};
\end{axis}

\begin{axis}[%
width=1.019in,
height=1.019in,
at={(0.795in,0.378in)},
scale only axis,
point meta min=-3,
point meta max=3,
axis on top,
xmin=0.5,
xmax=50.5,
xtick={\empty},
ymin=0.5,
ymax=50.5,
ytick={\empty},
axis background/.style={fill=white},
legend style={legend cell align=left, align=left, draw=white!15!black}
]
\addplot [forget plot] graphics [xmin=0.5, xmax=50.5, ymin=0.5, ymax=50.5] {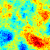};
\end{axis}

\begin{axis}[%
width=1.019in,
height=1.019in,
at={(2.343in,0.378in)},
scale only axis,
point meta min=-3,
point meta max=3,
axis on top,
xmin=0.5,
xmax=50.5,
xtick={\empty},
ymin=0.5,
ymax=50.5,
ytick={\empty},
axis background/.style={fill=white},
legend style={legend cell align=left, align=left, draw=white!15!black},
colormap/jet,
colorbar right,
colorbar style={anchor=south west, at={(2.716,0)}, height=1*\pgfkeysvalueof{/pgfplots/parent axis height}}
]
\addplot [forget plot] graphics [xmin=0.5, xmax=50.5, ymin=0.5, ymax=50.5] {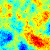};
\end{axis}

\begin{axis}[%
width=1.019in,
height=1.019in,
at={(3.89in,0.378in)},
scale only axis,
point meta min=-3,
point meta max=3,
axis on top,
xmin=0.5,
xmax=50.5,
xtick={\empty},
ymin=0.5,
ymax=50.5,
ytick={\empty},
axis background/.style={fill=white},
legend style={legend cell align=left, align=left, draw=white!15!black}
]
\addplot [forget plot] graphics [xmin=0.5, xmax=50.5, ymin=0.5, ymax=50.5] {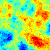};
\end{axis}
\end{tikzpicture}%
};
	\node [above right] at (3,3.7) {
%
%
\begin{tikzpicture}

\begin{axis}[%
width=1.019in,
height=1.019in,
at={(0.795in,1.935in)},
scale only axis,
point meta min=-3,
point meta max=3,
axis on top,
xmin=0.5,
xmax=50.5,
xtick={\empty},
ymin=0.5,
ymax=50.5,
ytick={\empty},
axis background/.style={fill=white},
legend style={legend cell align=left, align=left, draw=white!15!black}
]
\addplot [forget plot] graphics [xmin=0.5, xmax=50.5, ymin=0.5, ymax=50.5] {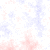};
\end{axis}

\begin{axis}[%
width=1.019in,
height=1.019in,
at={(2.343in,1.935in)},
scale only axis,
point meta min=-3,
point meta max=3,
axis on top,
xmin=0.5,
xmax=50.5,
xtick={\empty},
ymin=0.5,
ymax=50.5,
ytick={\empty},
axis background/.style={fill=white},
legend style={legend cell align=left, align=left, draw=white!15!black},
colormap={mymap}{[1pt] rgb(0pt)=(0,0,1); rgb(31pt)=(1,1,1); rgb(32pt)=(1,1,1); rgb(63pt)=(1,0,0)},
colorbar right,
colorbar style={anchor=south west, at={(2.716,0)}, height=1*\pgfkeysvalueof{/pgfplots/parent axis height}}
]
\addplot [forget plot] graphics [xmin=0.5, xmax=50.5, ymin=0.5, ymax=50.5] {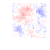};
\end{axis}

\begin{axis}[%
width=1.019in,
height=1.019in,
at={(3.89in,1.935in)},
scale only axis,
point meta min=-3,
point meta max=3,
axis on top,
xmin=0.5,
xmax=50.5,
xtick={\empty},
ymin=0.5,
ymax=50.5,
ytick={\empty},
axis background/.style={fill=white},
legend style={legend cell align=left, align=left, draw=white!15!black}
]
\addplot [forget plot] graphics [xmin=0.5, xmax=50.5, ymin=0.5, ymax=50.5] {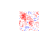};
\end{axis}
\end{tikzpicture}%
};
	
	\tikzmath{		\y1 = 4.5; \y2 = 8.4;   \y3 = 12.4;
		\x1 = 3.4; \x2 = 7; \x3 = -0.5;
		\alsx = 6.8;  
		\alsy = 9.8;  
		\arsx = 9.9;  
		\arsy = \alsy;  
		\atl = 2.4;
		\atb = 2.3;
	};
	\node[align=center] at (\y1,\x3) {\fontsize{10pt}{12pt}\selectfont pCN-MCMC};
	\node[align=center] at (\y2,\x3) {\fontsize{10pt}{12pt}\selectfont sequential pCN-MCMC};
	\node[align=center] at (\y3,\x3) {\fontsize{10pt}{12pt}\selectfont sequential Gibbs};
	
	
	\node[align=right,left] at (3,9.8) {\fontsize{10pt}{12pt}\selectfont Previous Sample \\ $\pai$};
	\node[align=right,left] at (3,5.2) {\fontsize{10pt}{12pt}\selectfont Change \\ $\pai - \pait$};
	\node[align=right,left] at (3,1.5) {\fontsize{10pt}{12pt}\selectfont Proposed Sample  \\ $\pait$};
	
	\node[align=center] at (\y1,\x2) {\fontsize{20pt}{12pt}\selectfont $+$ };
	\node[align=center] at (\y2,\x2) {\fontsize{20pt}{12pt}\selectfont $+$ };
	\node[align=center] at (\y3,\x2) {\fontsize{20pt}{12pt}\selectfont $+$ };

	\node[rotate=90,align=center] at (\y1+0.05,\x1) {\fontsize{20pt}{12pt}\selectfont $=$ };
	\node[rotate=90,align=center] at (\y2+0.05,\x1) {\fontsize{20pt}{12pt}\selectfont $=$ };
	\node[rotate=90,align=center] at (\y3+0.05,\x1) {\fontsize{20pt}{12pt}\selectfont $=$ };

	\draw [line width=0.5mm,->] (\alsx,\alsy) --  (\alsx-\atl,\alsy) -- (\alsx-\atl,\alsy-\atb);
	\draw [line width=0.5mm,->] (\arsx,\arsy) --  (\arsx+\atl,\arsy) -- (\arsx+\atl,\arsy-\atb);
	\draw [line width=0.5mm,->] (\y2-0.1,8.2) --  (\y2-0.1,\arsy-\atb) ;
	
	\end{tikzpicture}
	\caption{Proposal step of pCN-MCMC, \mymethod \ and sequential Gibbs sampling. The pCN-MCMC makes a small global change. The sequential Gibbs sampling makes a large local change. The \mymethod \  makes a medium change in a medium-sized area. This figure is only for visualization. Realistic problems lead to smaller changes in all three algorithms.}
	\label{fig_concept}
\end{figure}
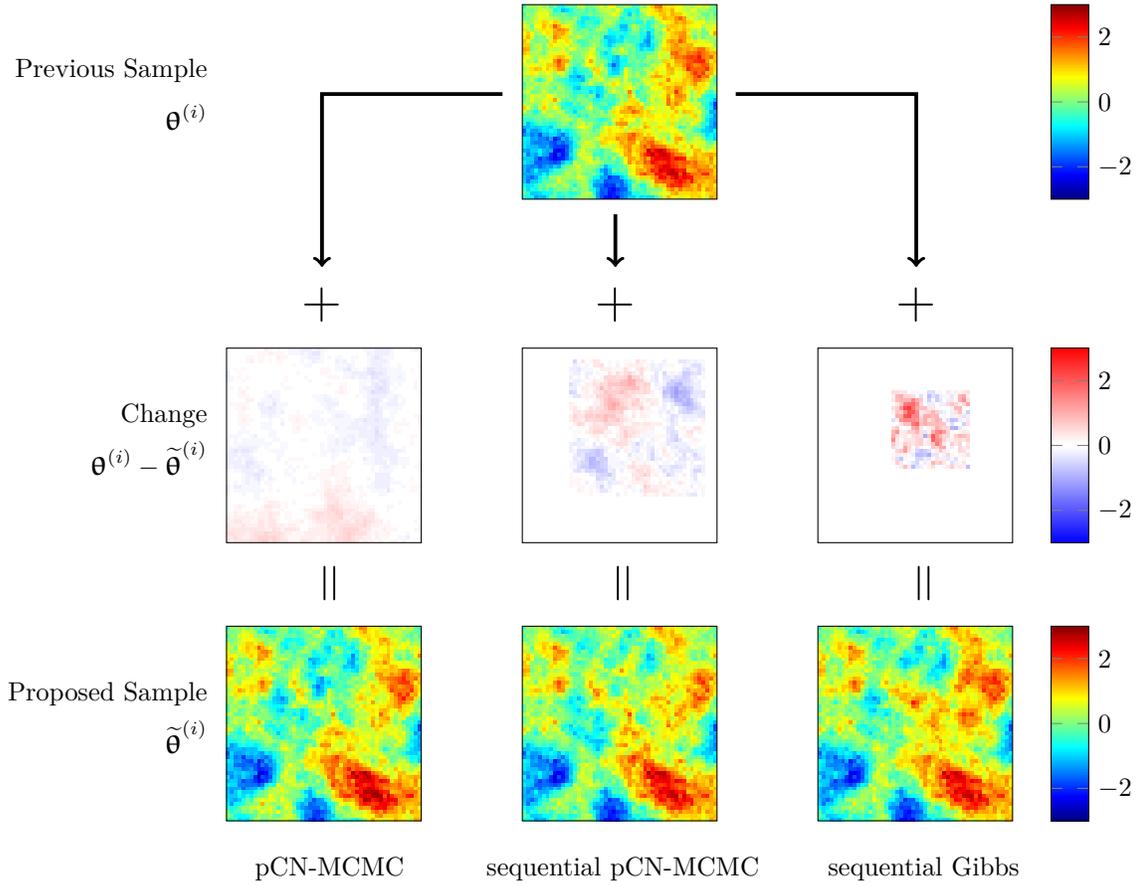

\noindent In this section, we present our proposed \mymethod .
To understand the underlying idea, let us look at the different MCMCs from a conceptual point of view. On the one hand, the proposal method of the pCN-MCMC makes global, yet small, changes that sample from the prior (left column of figure \ref{fig_concept}). On the other hand, the sequential Gibbs method makes local, yet large, changes that also sample from the prior (right column of figure \ref{fig_concept}). We want to combine these two approaches to make medium changes in a medium sized area which again sample from the prior (center column of figure \ref{fig_concept}).\\
We take the same preparatory steps as in the sequential Gibbs approach (equations \ref{eqn_M}-\ref{eqn_gibbs_covariance}). However, we propose a new sample within the resampling box based on the pCN approach (equation \ref{eqn_pcn})

\begin{align}
\label{eqn_proposal_sequential_pcn}
\paramt_1 = \sqrt{(1-\beta^2)}\left(\pai_1 - \widetilde{\boldsymbol{\upmu}}_1  \right) + \beta  \boldsymbol{\upxi} + \widetilde{\boldsymbol{\upmu}}_1 , \quad \boldsymbol{\upxi} \sim N(\boldsymbol{0},\widetilde{\boldsymbol{\Sigma}}_{11}) \ .
\end{align}

$\widetilde{\boldsymbol{\upmu}}_1$ and $\widetilde{\boldsymbol{\Sigma}}_{11}$ have been defined in equation \ref{eqn_gibbs_mu} and equation \ref{eqn_gibbs_covariance}, respectively. 
Consequently, the proposal distribution is defined as

\begin{align}
\paramt = \boldsymbol{\mathrm{M}}_\kappa^{T} \left[\begin{matrix}
\paramt_1 \\
\param_2 \\
\end{matrix}\right] , \quad \paramt_1 \sim \text{see equation \ref{eqn_proposal_sequential_pcn} }  \ .
\end{align}

This allows for sequential pCN-MCMC proposal in blocks of the parameter space. 

Any combination of the tuning-parameter $\beta$ of the pCN-MCMC approach and the tuning-parameter $\kappa$ of the Gibbs approach can be chosen. Alike the pure cases, increasing $\kappa$ and $\beta$ will lead to larger changes in subsequent samples and lower acceptance rates. 
Both, the sequential Gibbs and the pCN-MCMC are special cases of the proposed sequential pCN-MCMC. The sequential Gibbs method is the special case for $\beta = 1$ and $\kappa = 1$ leads to the pCN-MCMC approach.

\subsection{Adaptive sequential pCN-MCMC}
\label{chp_optimization}
Section \ref{chp_alpha} stated that the acceptance rate $\overline{\alpha}$ is often used to tune the tuning-parameters of MCMC methods. This tuning does only work for one tuning-parameter. In our proposed method, we have two tuning-parameter, namely the box size $\kappa$ of the sequential Gibbs method and the pCN parameter $\beta$. The presence of two tuning-parameters destroys the uniqueness of the optimum ($\overline{\alpha} = 0.234$). Hence, we need to find the optimal (or good enough) parameter combination in another way. 

We propose an adaptive version of the sequential pCN-MCMC, which finds the optimal tuning-parameter during its burn-in period. For this, we take a gradient descent approach. We evaluate the performance of tuning-parameters by running the MCMC for $N_{hp}$ steps, e.g. $10^3$ steps, with the same tuning-parameters. Then we evaluate the produced subsample based on the efficiency (see section \ref{chp_efficiency}). The efficiency is used because it can be evaluated, unlike the R-statistic or KL-divergence, during the runtime of the algorithm. The R-statistic and the KL-divergence need several independent chains to assess performance. The presented approach uses one chain, so it cannot build its optimization on the R-statistic or KL-divergence (while it can use the efficiency), but we use them as independent and complementary checks.

However, the autocorrelation, on which the efficiency is based on, is normalized by the total variance of the sample. Hence, the efficiency is independent of the total variance of the sample. To favor tuning-parameters that explore the posterior as much as possible, even in a small subsample, we add a standard derivation term to the objective function. For an increasing size of the subsamples, the effect of the standard derivation diminishes as the standard derivation of all subsamples converges to the standard derivation of the posterior. 
This leads to the objective function (which should be maximized) of a subsample 

\begin{align}
f(\beta,\kappa) =  \frac{1}{N_p} \sum_{j=1}^{N_p} eff_j \cdot s_j
\end{align}
where $s_j$ is the current standard derivation of the $j$-th parameter.

The adaptive sequential pCN-MCMC starts with a random (or expert-guess) tuning-parameter combination $\beta_0,\kappa_0$. Then, the derivatives of the objective functions are approximated numerically by 

\begin{align}
\frac{\partial f}{\partial \beta}  =  \frac{f(\beta_k^+,\kappa_k) -f(\beta_k^-,\kappa_k)}{\beta_k^+-\beta_k^-}
\end{align}
\begin{align}
\frac{\partial f}{\partial \kappa}  =  \frac{f(\beta_k,\kappa_k^+) -f(\beta_k,\kappa_k^-)}{\kappa_k^+-\kappa_k^-}
\end{align}
with
\begin{align}
\beta_k^+ = \beta_k \cdot \delta_\beta ,& \ \ 
\beta_k^- = \beta_k \cdot \frac{1}{\delta_\beta} \\ 
\kappa_k^+ = \kappa_k \cdot \delta_\beta ,& \ \ 
\kappa_k^- = \kappa_k \cdot \frac{1}{\delta_\beta} .
\end{align}
where $\delta_\beta = \delta_\kappa = \sqrt{2}$ in our implementation. This selection of evaluation points leads to an equal spacing of evaluation points in the log-space.  
Next, the algorithm moves a predefined distance (in the loglog-space) towards the steepest descent (here: rise, because we maximize the objective function) and restarts evaluating the tuning-parameters. 

Two things are important here. First, we use the loglog-space because the results (figure \ref{fig_results_1} and \ref{fig_results_2}) suggest that the efficiency does not have sudden jumps in the loglog-space which makes the optimization easy. Optimizing in the "normal" space would lead to a more complex optimization problem due to a more complex strucutre of good values (banana shaped instead of a straight line) and high derivatives for small $\kappa$ and $\beta$. Second, the predefined distance is important because the objective function is stochastic, i.e. starting it twice with the same parameters will not lead to the same result $f$. Not predefining the distance (which is done in vanilla steepest descent methods) leads to some, randomly happening, high derivatives that prevent convergence of the algorithm.

\section{Testing cases and implementation}
\label{chp_test_cases}
\subsection{Testing procedure}

We test our method by inferring the hydraulic conductivity of a confined aquifer based on measurements of hydraulic heads in a fully saturated, steady-state, 2D groundwater flow model. The data for inversion are generated synthetically with the same model as used for inversion. We are interested in several different cases: First, we test our method in a coarse-grid resolution $[50\times 50]$ cells. Here, we systematically test tuning-parameter ($\beta,\kappa$) combinations to find the optimal parameter combination. Further, we developed the adaptive sequential pCN-MCMC in this case. 

Second, we used the same reference solution with more informative measurement data and conducted the same systematic testing of tuning-parameters as in test case 1. We test the adaptive sequential pCN-MCMC on this new test case on which it was not developed. This enables us to make (more or less) general statements about the performance of this tool. After that, we try different variants of the original model to test our algorithm in different conditions and at higher resolutions.

In all test cases we run the sequential pCN-MCMC methods with $2$ million samples of which we save every $200$th sample. We discard the first half of each run due to burn-in and calculate the metrics presented in section \ref{chp_metrics} based on the second half. Hence, each metric is calculated using $5000$ samples. This is done three times for each tuning-parameter combination in test case $1 -4$ and we report the mean value in the following. In test case $5$, we test each adaptive method (adapitve sequential Gibbs, adaptive pCN-MCMC, adaptive sequential pCN-MCMC) three times and report the mean values of these runs. Here, the adapitve sequential Gibbs (or adaptive sequential pCN-MCMC) correspons to the case where we optimize $\kappa$ (or $\beta$) as proposed in section \ref{chp_optimization} and set $\beta = 1$ (or $\kappa = 1$).   

To evaluate the KL-divergence, a reference solution is calculated using the best tuning-parameter combination of each test case and running the sequential pCN-MCMC for $10$ million samples. We save every $200$th sample and remove the first million samples as burn-in.

\subsection{Description of test cases}

\subsubsection{Base case}
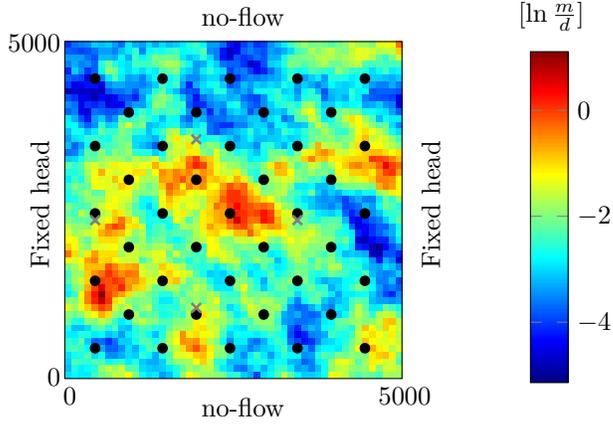
\begin{figure}
	\begin{tikzpicture}
	\node [above right] at (0.5,0.5) {
%
%
\definecolor{mycolor1}{rgb}{0.00000,0.44700,0.74100}%
\begin{tikzpicture}

\begin{axis}[%
width=1.765in,
height=1.765in,
at={(0.268in,0.238in)},
scale only axis,
point meta min=-5.14027527512201,
point meta max=1.10532523127064,
axis on top,
xmin=0,
xmax=5000,
xtick={\empty},
ymin=0,
ymax=5000,
ytick={\empty},
axis background/.style={fill=white},
colormap/jet,
colorbar right,
colorbar style={anchor=south west, at={(1.381,-0.012)}, height=0.982*\pgfkeysvalueof{/pgfplots/parent axis height}}
]
\addplot [forget plot] graphics [xmin=0, xmax=5000, ymin=0, ymax=5000] {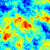};
\addplot [color=black, line width=2.0pt, draw=none, mark size=1.0pt, mark=*, mark options={solid, fill=white, black}, forget plot]
  table[row sep=crcr]{%
450	450\\
450	1450\\
450	2450\\
450	3450\\
450	4450\\
1450	450\\
1450	1450\\
1450	2450\\
1450	3450\\
1450	4450\\
2450	450\\
2450	1450\\
2450	2450\\
2450	3450\\
2450	4450\\
3450	450\\
3450	1450\\
3450	2450\\
3450	3450\\
3450	4450\\
4450	450\\
4450	1450\\
4450	2450\\
4450	3450\\
4450	4450\\
950	950\\
950	1950\\
950	2950\\
950	3950\\
1950	950\\
1950	1950\\
1950	2950\\
1950	3950\\
2950	950\\
2950	1950\\
2950	2950\\
2950	3950\\
3950	950\\
3950	1950\\
3950	2950\\
3950	3950\\
};
\addplot [color=mycolor1, line width=1.0pt, draw=none, mark size=2.5pt, mark=x, mark options={solid, gray}, forget plot]
  table[row sep=crcr]{%
450	2350\\
3450	2350\\
1950	3550\\
1950	1050\\
};
\end{axis}
\end{tikzpicture}%
};
	\node [below] at (0.7,0.7) {0};
	\node [left] at (0.7,0.7) {0};
	\node [left] at (0.7,5.15) {5000};
	\node [below] at (5.15,0.7) {5000};
	\node [rotate=0] at (7.1,5.5) {\text{[$\ln\frac{m}{d}$]}};
	\node at (3,5.5) {no-flow};
	\node at (3,0.3) {no-flow};
	\node [rotate=90] at (0.3,3) {Fixed head};
	\node [rotate=90] at (5.5,3) {Fixed head};
	\end{tikzpicture}
	\caption{Log-conductivity field of the synthetic reference aquifer with a fixed head boundary condition on the left side and right side, no-flow boundary conditions at top and bottom and groundwater extraction wells marked with grey crosses. The positions of the measurement wells are marked in black.}
	\label{fig_setup}
\end{figure}

We consider an artificial steady-state groundwater flow in a confined aquifer test case as proposed in \cite{Xu2020}.
It has a size of $5000 \times 5000[m]$ and is discretized into $50 \times 50$ cells as shown in figure \ref{fig_setup}.

We assume a multi-Gaussian prior model with mean  $-2.5 [\frac{m}{d}]$ and variance equal to $1$. Further, we assume an anisotropic exponential variogram with lengthscale parameters $[1500,2000]$ rotated by 135 degrees. The higher value of the lengthscale is pointing from the bottom left to the top right (see figure \ref{fig_setup}). 

We assume no-flow boundary conditions at the top and bottom boundary, a fixed head boundary condition with $h=20m$ at the left and $h=0m$ at the right side. Further, we assume four groundwater extraction wells as shown in table \ref{tbl_wells}.

\begin{table}[]
	\centering
	\begin{tabular}{c|c|c}
		position x [$m$]   &  position y  [$m$]  &  pump strength [$\frac{m^3}{d}$]    \\
		\hline
		500  & 2350 & 120  \\
		3500 & 2350 & 70   \\
		2000 & 3550 & 90   \\
		2000 & 1050 & 90 
	\end{tabular}
	\caption{Position and pumping strength of groundwater extraction wells.}
	\label{tbl_wells}
\end{table}

Figure \ref{fig_setup} shows the hydraulic conductivity distribution of the artificial true aquifer and the $41$ measurement locations marked in black. 
We corrupt each of the $41$ simulated (with the hydraulic conductivity of the artificial true aquifer) head values with a variate drawn at random from a zero-mean normal distribution with variance of $0.05 [m]$ to obtain synthetic data for the inversion

The flow in the domain can be described by the saturated groundwater flow equation 
\begin{align}
\nabla \cdot \left[ K(x,y)\nabla h(x,y,t) \right] = \eta(x,y),
\end{align}
where $K$ is the isotropic hyrdaulic conductivity, $h$ is the hydraulic head and $\eta$ encapsulates all source and sink terms.
We solve the equation using the flow solver described in \citeA{Nowak2005Diss} and \citeA{Nowak2008} numerically.

\subsubsection{Test case 2}
The sole difference between test case 2 and the base case is that a standard derivation of the measurement error of $0.02 [m]$ is used. This leads to a more likelihood-dominated Bayesian inverse problem. Hence, slightly changing parameters results in higher differences in the corresponding likelihoods. This leads to a smaller acceptance rate which is dependent on the quotient of subsequent likelihoods. To keep a constant acceptance rate, a smaller jump width is needed. Hence, we expect the optimal $\kappa$ and $\beta$ to decrease for a more likelihood dominated posterior.  

\subsubsection{Test case 3}
The only difference between test case 3 and the base case is that only $16$ instead of $41$ measurement positions were used. This leads to a less likelihood-dominated Bayesian inverse problem and reverses the variation done in test case 2.

\subsubsection{Test case 4}
Test case 4 has different reference solution with a Matern covariance with $\nu = 2.5$ and isotropic lengthscale parameter $\lambda=1000$. All other parameters are identical to the base case setup. We do this variation to test the influence of the prior covariance structure on our proposed method. 

\subsubsection{Test case 5}
Test case 5 uses a refined reference solution with a $[100,100]$ discretization grid. Here, the higher discretization in the inversion makes the problem numerically more expensive, serving to demonstrate the efficiency and applicability of our method.

\subsubsection{Test case 6}
Test case 6 uses the artificial true aquifer of the base case. However, instead of infering the hydraulic conductivity with head measurements, we infer the hydraulic conductivity with concentration measurements. 
We assume a dissolved, conservative, nonsorbing tracer transported by the advection dispersion equation in steady-state flow

\begin{align}
\label{eqn_transport}
\nabla\cdot( \boldsymbol{v} c - \boldsymbol{D}\nabla c) = 0 
\end{align}
with the seepage velocity $\boldsymbol{v}$ and the local dispersion tensor $\boldsymbol{D}$.
We assume that some concentration $c$ is constantly entered in the center third of the left boundary by setting the boundary condition to $\frac{c}{c_0} = 1$ there, where $c_0$ is e.g., a solubility limit.
Then, we calculate the steady-state solution of equation \ref{eqn_transport} under time-constant boundary conditions and measure the concentration at the 5 upmost right measurement locations shown in figure \ref{fig_setup}. Here, we assume a standard derivation of the measurement error equal to $0.05$. This test case shows the influence that different measurements types have to our results.  
We use the flow and transport solver described in \citeA{Nowak2005Diss} and \citeA{Nowak2008} to solve the equations numerically.

\section{Results}
\label{chp_results}
\subsection{Base case}
\label{chp_test_case_1}
\begin{figure}
	\centering
	\input{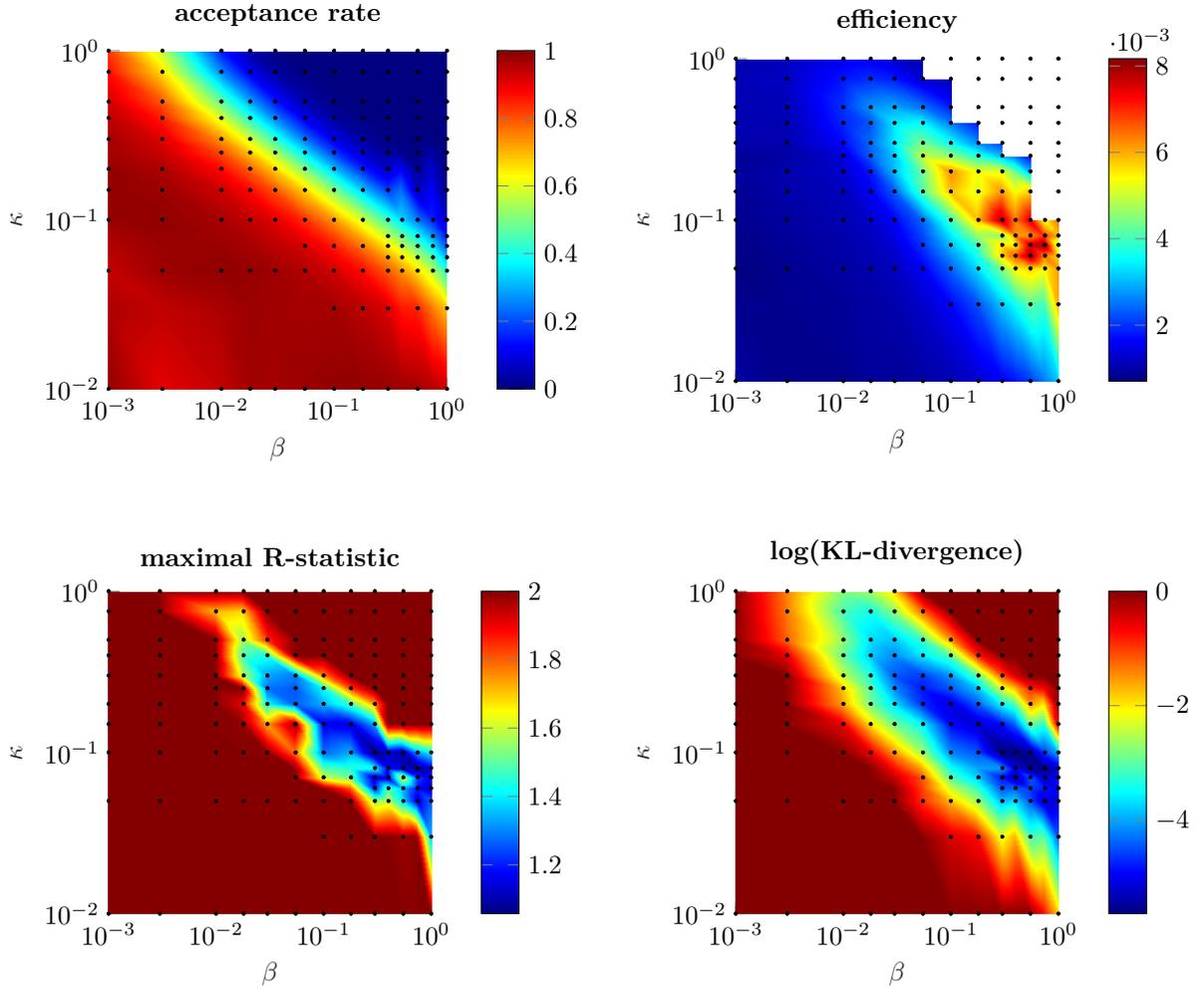}
	\caption{Results of base case. The sequential pCN-MCMC was tested with all combinations marked by black points. All in-between values are interpolated. The efficiency was not calculated in the top right corner (white area) because the MCMC did not finish the burn-in during runtime. The pCN-special case is at the top with $\kappa = 1$ and the Gibbs special case is at the right with $\beta = 1$. The acceptance rate shows no unique optimum ($\alpha = 0.234$), but rather a line of optimal tuning-parameters combinations. The efficiency, the R-statistic and the KL divergence indicate similar optima. The optimal efficiency is at $\beta = 0.75$ and $\kappa = 0.07$. }
	\label{fig_results_1}
\end{figure}

The acceptance rate of the MCMC, the efficiency, the R-statistic and the (log of the) KL divergence to a reference solution are visualized in figure \ref{fig_results_1}. As discussed in section \ref{chp_metrics}, we aim for an R-statistic equals $1$,  a high efficiency and a low log KL conductivity which corresponds to an acceptance rate equals $23\%$.  We focus on two things in this plot.
First, we see that all metrics have a similar appearance. Hence, a tuning-parameter combination that performs well in one metric also performs well in the other two metrics and vice versa. Second, the optimal tuning-parameter combination ($\beta, \kappa$) is arround $\beta = 0.75$ and $\kappa = 0.07$ for all norms. Hence, neither the pCN ($\kappa=1$) nor the sequential Gibbs special case ($\beta=1$) is optimal. The sequential-pCN MCMC has a better performance than the special cases. Further, the efficiency plot (or table \ref{tbl_results_cases}) indicates a speedup of approximately $5.1$ over pCN-MCMC and of $1.3$ over sequential Gibbs sampling.

\subsection{Test case 2}
\label{chp_test_case_2}
\begin{figure}
	\centering
	\input{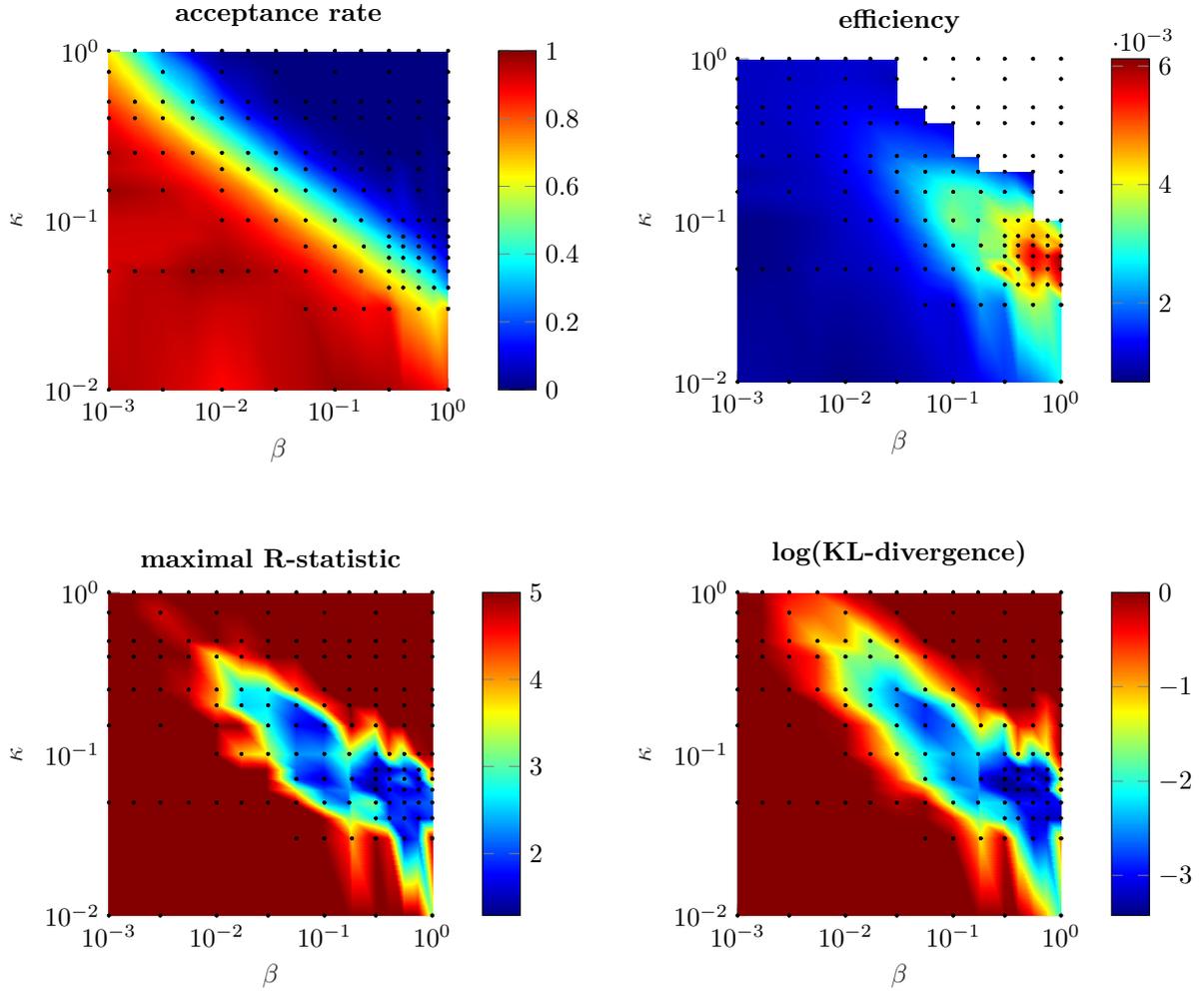}
	\caption{Results of test case 2. The sequential pCN-MCMC was tested with all combinations marked by black points. All in-between values are interpolated. The efficiency was not calculated in the top right corner (white area) because the MCMC did not finish the burn-in during runtime. The pCN-special case is at the top with $\kappa = 1$ and the Gibbs special case is at the right with $\beta = 1$. The acceptance rate shows no unique optimum ($\alpha = 0.234$), but rather a line of optimal tuning-parameters combinations. The efficiency, the R-statistic and the KL divergence indicate similar optima. The optimal efficiency is at $\beta = 1$ and $\kappa = 0.06$.}
	\label{fig_results_2}
\end{figure}

Using the second test case, we visualize the acceptance rate of the MCMC, the efficiency, the R-statistic and the KL divergence to a reference solution in figure \ref{fig_results_2}.
As discussed in section \ref{chp_test_case_2}, we expect smaller optimal tuning-parameters compared to test case 1. Comparing figure \ref{fig_results_1} and figure \ref{fig_results_2}, we see that our expectations are partly met. The light blue line of acceptance rates of approx $40 \%$ is moving to the bottom left, to smaller $\kappa$ and $\beta$, as expected. However, the optimal values change from $\beta = 0.75$ and $\kappa = 0.07$ in the first testcase to  $\beta = 1$ and $\kappa = 0.06$ in the second case. 
Hence, the optimal tuning-parameters tend more towards the Gibbs approach (no pCN correlation at $\beta=1$, instead a smaller window). In fact, we find that the sequential Gibbs approach is as good as the sequential pCN-MCMC approach because the optimal parameter $\beta$ equals 1.

\subsection{Further test cases}

Next, we test our algorithm with fewer measurement locations (case 3), with a Matern variogram model (case 4) and with a finer discretization (case 5) as summarized in table \ref{tbl_results_cases}. In short, the results indicate that the sequential-pCN MCMC has at least the same performance as the Gibbs or pCN-MCMC approach. We achieve a speedup (measured by the ratio of efficiencies) of $1-5.5$ over pCN and of $1-6.5$ over Gibbs.

In test cases 1-4, structured testing as shown in section \ref{chp_test_case_1} and section \ref{chp_test_case_2} was performed. For the high-dimensional test case 5 and the transport test case 6 this testing procedure was too computationally expensive. Hence, the previous tested adaptive sequential pCN-MCMC was used to find the best parameter distribution.

\begin{table}[]
	\begin{tabular}{l|ccc|ccc|ccc|cc}
		& \multicolumn{3}{|c|}{Efficiency}  & \multicolumn{3}{|c|}{maximal R-statistic}  & \multicolumn{3}{|c|}{KL-divergence} & \multicolumn{2}{|c}{optimal} \\
		& se. pCN & pCN & Gibbs& se. pCN & pCN & Gibbs& se. pCN & pCN & Gibbs  & $\beta$  & $\kappa$ \\ \hline 		
		base case & 0.0082 & 0.0016 & 0.0065 & 1.0572 & 2.0697 & 1.1222 & 0.0036 & 0.0660 & 0.0066 &  0.75 & 0.07 \\
		case 2 & 0.0061 & 0.0011 & 0.0061 & 1.2888 & 5.4102 & 1.9264 & 0.0326 & 0.5923 & 0.0419 & 1 & 0.06 \\
		case 3 & 0.0087 & 0.0029 & 0.0063 & 1.0521 & 1.2661 & 1.0854 &  0.0051 & 0.0212 & 0.0072 &0.3 & 0.1  \\
		case 4 & 0.0019 & 0.0012 & 0.0018 & 1.8374 & 3.9433 & 2.5057 &  0.0694 & 0.6710 & 0.1103 &0.3 & 0.1 \\
		case 5* & 0.0065  &  0.0019 & 0.0065  & 1.1315  &  2.25  & 1.1315     &  --- &   --- &   ---   & 1 & 0.0579 \\
		case 6* & 0.1711  &  0.1711 & 0.0264  & 1.002  &  1.002  & 1.024     &  --- &   --- &   ---   & 0.1741 & 1
	\end{tabular}
	\caption{Test metrics of algorithms with optimal tuning-parameters. Further, the optimal value of the sequential pCN-MCMC tuning-parameters with respect to the efficiency is presented. *:No systematic testing was performed. Optimal tuning-parameters were found using the adaptive sequential pcn-MCMC.}
	\label{tbl_results_cases}
\end{table}

\subsection{Adaptive sequential pCN-MCMC}

\begin{figure}
	\centering
	\input{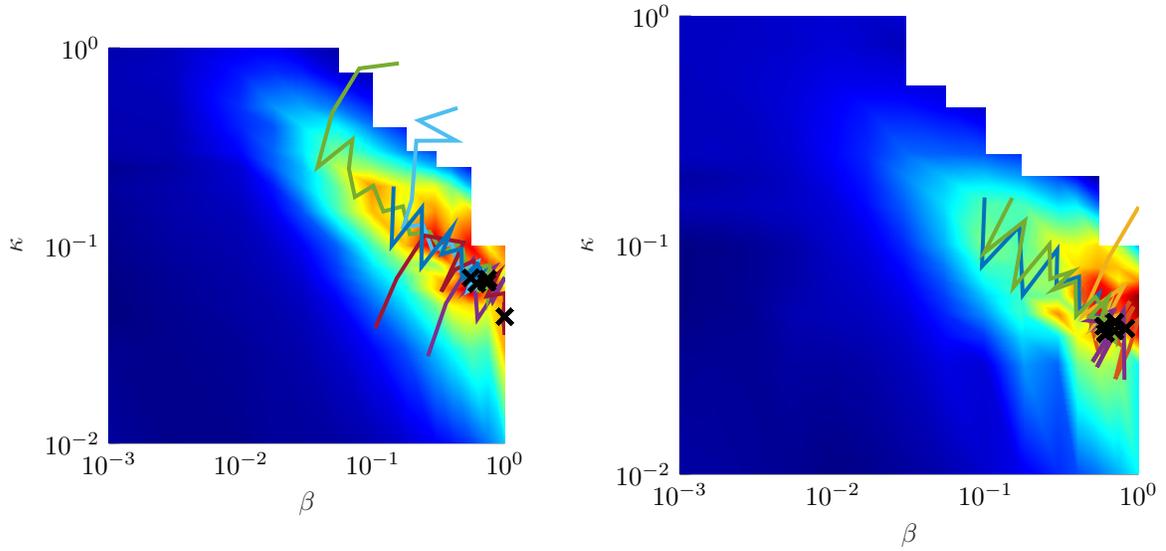}
	\caption{Convergence of adaptive sequential pCN MCMC in test case one (left) and test case two (right). The chosen parameter for production is marked with a cross. The efficiency is shown in the background.}
	\label{fig_optimization}
\end{figure}

\citeA{Gelman1996} stated that a wide range of tuning-parameters are satisfactory (close to optimal) for MCMCs with one tuning-parameter. Figure \ref{fig_results_1} shows that this holds for this two-tuning-parameter MCMC as well. The efficiency, the R statistic and the KL divergence have a broad area of near-optimal tuning-parameters. Hence, the adaptive sequential pCN-MCMC only needs to find some point in this good area to choose near-optimal parameters. Figure \ref{fig_optimization} shows 5 paths of the adaptive sequential pCN-MCMC during burn-in with random start tuning-parameters for the first and second test cases. Each path consists of all tested tuning-parameter combinations. Figure  \ref{fig_optimization} shows that all paths converge to the targeted area of near-optimal tuning-parameters. 

We note here, that many tuning-parameter optimization steps $N_{hp}$ are needed to achieve these results. Using fewer MCMC steps per tuning-parameter iteration leads to less exact tuning-parameter tuning. 
Although the tuning is less exact with smaller $N_{hp}$, we find in additional experiments that the adaptive sequential pCN-MCMC converges to acceptable tuning-parameters with small $N_{hp}$ due to broad high-efficiency areas.
The needed size of $N_{hp}$ depends on the amount of measurement information. Having more information leads, even for near-optimal tuning-parameters, to a smaller step size (of the MCMC) and lower efficiency. Hence, we need more samples for good MCMC results and simultaneously for each tuning-parameter tuning iteration.

\section{Discussion}
\label{chp_discussion}
\subsection{Global versus local proposal steps}
The results in table \ref{tbl_results_cases} indicate that the pure (local)  Gibbs approach is superior to the pure (global) pCN approach when used with head measurements. We did further testing using a simple Kriging (measuring the parameters directly) example and found that the Gibbs approach is superior to the pCN approach in that case as well.
In transport scenarios (test case 6), the (global) pCN approach is superior to the (local) Gibbs approach. 

We explain this behavior in the following way: Direct (and head measurements) typically yield us local (or relatively local) information of the aquifer. It only lets us infer the hydraulic conductivity in a small area around the measurement location because the influence of hydraulic conductivity on the measurement decreases rapidly with distance.
If we measure transport (concentration), almost all parameters in the domain have similar influence on the measurement.  Visual examples of the corresponding cross-correlation between heads, concentration and hydraulic conductivity are provided in e.g. \cite{Nowak2008} and \cite{Schwede2008}.
This leads to the conclusion, that measurements with localized information (head, direct measurements) work better with local updating schemes (Gibbs), whereas measurements with global information (transport) work better with global updating schemes (pCN). 
Note, that we only tested these algorithms on a few geostatistical problems and encourage researchers to compare global and local proposal steps and endorse or oppose our findings.

\subsection{Limits of sequential pCN-MCMC}

\noindent In our test cases, the sequential pCN-MCMC and the sequential Gibbs approach have higher efficiencies than the pCN-MCMC. 
However, this speedup comes at the increased cost of the proposal step. Computing the conditional probability (equation \ref{eqn_gibbs_mu} and \ref{eqn_gibbs_covariance}) is time consuming due to the computation of $\boldsymbol{\Sigma}_{22}^{-1}$. In most applications, the forward simulation (i.e. the calculation of the likelihood) is much more expensive than the inversion of the matrix $\boldsymbol{\Sigma}_{22}$ and the time difference in the proposal step can be neglected. However, with simple forwards problems, this might make the pCN approach a viable alternative to the sequential pCN-MCMC because all norms discussed in this paper neglect this time difference.

\citeA{Fu2008} discuss different schemes on how the conditional sampling can be performed without the need to compute $\boldsymbol{\Sigma}_{22}^{-1}$. The downside of these schemes is, that they do not sample from $N(\widetilde{\boldsymbol{\upmu}}_1,\widetilde{\boldsymbol{\Sigma}}_{11})$ directly but from some approximation of it. On regular, equispaced grids, FFT-related methods and sparse linear algebra methods \cite{Fritz2009,Nowak2013} offer exact and very fast solutions.

The sequential pCN-MCMC is not designed to handle multi-modal posteriors. However, applying parallel tempering approaches \cite<e.g.>{Laloy2016} can solve this challenge. They can be applied straightforward to our method but come with one downside. The adaptive sequential pCN-MCMC will not work in a parallel tempering setup because the efficiency depends on the autocorrelation. In parallel tempering, the autocorrelation is dominated by between-chain swaps and hence will not be a good estimator for the performance of MCMC. Finding another way to tune the tuning-parameters during burn-in will be the big challenge in generalizing the sequential pCN-MCMC to parallel tempering.

\subsection{Limits of optimizing the acceptance rate}
\noindent Our results suggest that acceptance rate values of $10\%$ - $60\%$, (case 1: $37\%$, case 2: $10\%$, case 3: $59\%$, case 4: $57\%$) are optimal. This is in conflict with the literature, especially \citeA{Gelman1996}, stating that acceptance rates equal $23.4\%$ are optimal for multi-Gaussian settings. The reason for this is the synthetic setting of \citeA{Gelman1996}. As a consequence, when using the acceptance rate $\alpha$ for optimizing the jump width, researchers should be aware that $\alpha = 23.4\%$ is not always optimal. 
Apart from that, we endorse \citeA{Gelman1996} that a wide area of acceptance rates lead to near-optimal results. Hence, the error by tuning for the wrong acceptance rate might be neglectable. 
We can not give a solution to this challenge but only point out that the suggested optimal acceptance rate of $23.4\%$ might not be the one you should always aim for.

\subsection{Transfer to multi-point geostatistics}
\noindent  The idea of building a hybrid between global and local jumps in parameter distributions can be applied to training image-based sampling methods in multi-point geostatistics as well. Both global \cite<resampling a percentage of  parameters scattered over the domain, e.g.>{Mariethoz2010a} and local approaches \cite<resampling a box of parameters, e.g.>{Hansen2012} exist and a combination might speed up the convergence for training image-based approaches as well. Thereby, a hybrid method should resample a higher percentage of scattered parameters in a larger box. 

\section{Conclusion}
\label{chp_conclusion}
We presented the sequential pCN-MCMC approach, a combination of the sequential Gibbs and the pCN-MCMC approach. Our approach has two tuning-parameters, the parameter $\beta$ of the pCN approach and $\kappa$ of the Gibbs approach. Setting either one of them to $1$ makes the algorithm collapse to either the pCN or the Gibbs approach. We show that the proposed method is more efficient than the sequential Gibbs and pCN-MCMC methods by testing all possible tuning-parameters of the sequential pCN-MCMC method.

Using more than one tuning-parameter has the downside that finding the optimal tuning-parameters is difficult. We presented the adaptive sequential pCN-MCMC to find good tuning-parameters during the burn-in of the algorithm.
This work can be extended to parallel tempering easily. However, the presented approach of finding the optimal tuning-parameters during burn-in needs to be adapted to fit the challenges of multiple chains.

\section*{Acknowledgement}
The implementation of the adaptive \mymethod \ is available at \href{https://bitbucket.org/Reuschen/sequential-pcn-mcmc}{https://bitbucket.org/Reuschen/sequential-pcn-mcmc}.
The authors thank Niklas Linde who was a valuable discussion partner with great ideas regarding the transfer to multi-point geostatistics and Sinan Xiao for revising the Manuscript. We thank the three anonymous reviewers for their constructive comments.
Funded by the Deutsche Forschungsgemeinschaft (DFG, German Research Foundation) – Project Number 327154368 – SFB 1313.
\section{Appendix}
The pseudo-codes of the algorithms discussed in this work are shown in the following. 


\begin{algorithm}[]
	\SetAlgoLined
	\SetKwInOut{Parameter}{Parameter}
	\SetKwInOut{Input}{Input}
	\SetKwInOut{Output}{Output}
	\Input{Prior probability density function $P(\param), \param \sim N(\boldsymbol{\upmu},\boldsymbol{\Sigma)}$ \\
		Likelihood function $L(\param)$}
	\Output{Samples $\param^{(i)}$ from posterior distribution $\pi(\param) = P(\param) L(\boldsymbol{\theta})$}
	\Parameter{$\kappa \in (0,1]$ }
	Set $i=0$ \\
	Draw $\pazero \sim N(\boldsymbol{\upmu},\boldsymbol{\Sigma})$ from prior \\
	\While{true}{
		$[\boldsymbol{\mathrm{M}}^{(i)},q^{(i)}]$ = GetBoxAlgorithm($\kappa$) \\
		$\left[
		\begin{array}{c}
		\param_1^{(i)}\\
		\param_2^{(i)}\\
		\end{array}
		\right]  =
		\boldsymbol{\mathrm{M}}^{(i)} \param^{(i)}$,
		$\left[
		\begin{array}{c}
		\boldsymbol{\upmu}_1\\
		\boldsymbol{\upmu}_2\\
		\end{array}
		\right]  =
		\boldsymbol{\mathrm{M}}^{(i)} \boldsymbol{\upmu}$,\\
		$\left[\begin{array}{cc}
		\boldsymbol{\Sigma}_{11} & \boldsymbol{\Sigma}_{12} \\ \boldsymbol{\Sigma}_{21} & \boldsymbol{\Sigma}_{22}
		\end{array}\right]  =
		\boldsymbol{\mathrm{M}}^{(i)} \boldsymbol{\Sigma} {\boldsymbol{\mathrm{M}}^{(i)}}^T $ \\ 
		
		$\widetilde{\boldsymbol{\upmu}}_1 = \boldsymbol{\upmu}_1 + \boldsymbol{\Sigma}_{12}\boldsymbol{\Sigma}_{22}^{-1}(\param_2^{(i)} -\boldsymbol{\upmu}_2)$ \\
		
		$\widetilde{\boldsymbol{\Sigma}}_{11}= \boldsymbol{\Sigma}_{11} - \boldsymbol{\Sigma}_{12}\boldsymbol{\Sigma}_{22}^{-1}\boldsymbol{\Sigma}_{21}$\\
		
		Propose $\boldsymbol{\widetilde{\theta}}_1^{(i)}  \sim N(\widetilde{\boldsymbol{\upmu}}_{1},\widetilde{\boldsymbol{\Sigma}}_{11})$ \\
		Set $\pait = {    \boldsymbol{\mathrm{M}}^{(i)}   }^T\left(
		\begin{array}{c}
		\pait_1\\
		\pai_2\\
		\end{array}
		\right) $\\
		Compute $\alpha(\pai,\pait) =  min \left[\frac{L(\pait)}{L(\pai)} ,1    \right]$ \\
		Draw $r \sim U(0,1)$\\
		\eIf{$r\leq\alpha$ \text{\normalfont (with probability} $\alpha$\text{\normalfont)}}{
			$\paii = \pait$
		}{
			$\paii = \pai$
		}
		$i=i+1$
	}
	\caption{sequential Gibbs sampling}
	\label{alg_gibbs}
\end{algorithm}

\begin{algorithm}[]
	\label{alg_box}
	\SetAlgoLined
	\SetKwInOut{Parameter}{Parameter}
	\SetKwInOut{Input}{Input}
	\SetKwInOut{Output}{Output}
	\SetKwInOut{Notation}{Notation}
	\Input{vectors of grid points positions $\boldsymbol{x},\boldsymbol{y}$\\
		lengths of domain $l_x,l_y$\\
		number of grid points $N_p$\\}
	\Output{permutation matrix $\boldsymbol{\mathrm{M}}$\\
		$q$, the size of all vectors with subscript $1$ (e.g. $\param_1$ )}
	\Parameter{$\kappa \in (0,1]$}
	\Notation{$\boldsymbol{x}(k)$ is the k-th element of vector $\boldsymbol{x}$\\}
	Draw midpoint $x^*,y^* \sim U(0,1)$\\
	$\boldsymbol{\mathrm{M}}=\boldsymbol{0}$ (of size $N_p \times N_p$)\\
	$q=0$,	$r=0$, $k=0$\\
	\While{$k < N_P$}{
		$k=k+1$\\
		\eIf{$\left|\frac{\boldsymbol{x}(k)}{l_x}-x^*\right|\leq\kappa \text{ and } \left|\frac{\boldsymbol{y}(k)}{l_y}-y^*\right|\leq\kappa$}{
			$q=q+1$\\
			$\boldsymbol{\mathrm{M}}(q,k)=1$\\
		}{
			$r=r+1$\\
			$\boldsymbol{\mathrm{M}}(N_p-r+1,k)=1$\\
		}
		
	}
	
	\caption{get box algorithm }
	
\end{algorithm}

\begin{algorithm}[]
	\SetAlgoLined
	\SetKwInOut{Parameter}{Parameter}
	\SetKwInOut{Input}{Input}
	\SetKwInOut{Output}{Output}
	\Input{Prior probability density function $P(\param), \param \sim N(\boldsymbol{\upmu},\boldsymbol{\Sigma})$ \\
		Likelihood function $L(\param)$}
	\Output{Samples $\param^{(i)}$ from posterior distribution $\pi(\param) = P(\param) L(\boldsymbol{\theta})$}
	\Parameter{$\beta\in (0,1]$ }
	Set $i=0$ \\
	Draw $\pazero \sim N(\boldsymbol{\upmu},\boldsymbol{\Sigma})$ from prior \\
	\While{true}{
		Propose $\boldsymbol{\widetilde{\theta}}^{(i)} = \sqrt{(1-\beta^2)}\left(\pai -  \boldsymbol{\upmu} \right) + \beta  \xi^{(i)} + \boldsymbol{\upmu} , \xi^{(i)} \sim N(0,{\boldsymbol{\Sigma}})$ \\
		Compute $\alpha(\pai,\pait) =  min \left[\frac{L(\pait)}{L(\pai)} ,1    \right]$ \\
		Draw $r \sim U(0,1)$\\
		\eIf{$r\leq\alpha$ \text{\normalfont (with probability} $\alpha$\text{\normalfont)}}{
			$\paii = \pait$
		}{
			$\paii = \pai$
		}
		$i=i+1$
	}
	\caption{pCN-MCMC}
	\label{alg_pcn}
\end{algorithm}

\begin{algorithm}[]
	\SetAlgoLined
	\SetKwInOut{Parameter}{Parameter}
	\SetKwInOut{Input}{Input}
	\SetKwInOut{Output}{Output}
	\Input{Prior probability density function $P(\param), \param \sim N(\boldsymbol{\upmu},\boldsymbol{\Sigma})$ \\
		Likelihood function $L(\param)$}
	\Output{Samples $\param^{(i)}$ from posterior distribution $\pi(\param) = P(\param) L(\boldsymbol{\theta})$}
	\Parameter{$\kappa \in (0,1]$\\
		$\beta\in (0,1]$ }
	Set $i=0$ \\
	Draw $\pazero \sim N(\boldsymbol{\upmu},\boldsymbol{\Sigma})$ from prior \\
	\While{true}{
		$[\boldsymbol{\mathrm{M}}^{(i)},q^{(i)}]$ = GetBoxAlgorithm($\kappa$) \\
		$\left[
		\begin{array}{c}
		\param_1^{(i)}\\
		\param_2^{(i)}\\
		\end{array}
		\right]  =
		\boldsymbol{\mathrm{M}}^{(i)} \param^{(i)}$, 		
		$\left[
		\begin{array}{c}
		\boldsymbol{\upmu}_1\\
		\boldsymbol{\upmu}_2\\
		\end{array}
		\right]  =
		\boldsymbol{\mathrm{M}}^{(i)} \boldsymbol{\upmu}$,	\\	
		$\left[\begin{array}{cc}
		\boldsymbol{\Sigma}_{11} & \boldsymbol{\Sigma}_{12} \\ \boldsymbol{\Sigma}_{21} & \boldsymbol{\Sigma}_{22}
		\end{array}\right]  =
		\boldsymbol{\mathrm{M}}^{(i)} \boldsymbol{\Sigma} {\boldsymbol{\mathrm{M}}^{(i)}}^T $\\ 
		
		$\widetilde{\boldsymbol{\upmu}}_1 = \boldsymbol{\upmu}_1 + \boldsymbol{\Sigma}_{12}\boldsymbol{\Sigma}_{22}^{-1}(\param_2^{(i)} -\boldsymbol{\upmu}_2)$ \\
		
		$\widetilde{\boldsymbol{\Sigma}}_{11}= \boldsymbol{\Sigma}_{11} - \boldsymbol{\Sigma}_{12}\boldsymbol{\Sigma}_{22}^{-1}\boldsymbol{\Sigma}_{21}$\\
		
		Propose $\boldsymbol{\widetilde{\theta}}_1^{(i)} = \sqrt{(1-\beta^2)}\left(\pai_1 - \widetilde{\boldsymbol{\upmu}}_1 \right) + \beta  \xi^{(i)} + \widetilde{\boldsymbol{\upmu}}_1 , \xi^{(i)} \sim N(0,\widetilde{\boldsymbol{\Sigma}}_{11})$ \\
		Set $\pait = {    \boldsymbol{\mathrm{M}}^{(i)}   }^T\left(
		\begin{array}{c}
		\pait_1\\
		\pai_2\\
		\end{array}
		\right) $\\
		Compute $\alpha(\pai,\pait) =  min \left[\frac{L(\pait)}{L(\pai)} ,1    \right]$ \\
		Draw $r \sim U(0,1)$\\
		\eIf{$r\leq\alpha$ \text{\normalfont (with probability} $\alpha$\text{\normalfont)}}{
			$\paii = \pait$
		}{
			$\paii = \pai$
		}
		$i=i+1$
	}
	\caption{\mymethod}
	\label{alg_block_pcn}
\end{algorithm}

\newpage
\bibliography{../library}

%
%
%
%
%

\end{document}